\documentclass{aastex62}

\usepackage{amsmath}
\usepackage{makecell}

\accepted{ApJ}

\shorttitle{GRB160625B Jet Structure}
\shortauthors{Strausbaugh et al.}

\begin{document}

\title{Evidence for a Bright-Edged Jet in the Optical/NIR Afterglow of GRB~160625B}

\correspondingauthor{Robert Strausbaugh}
\email{Robert.Strausbaugh@asu.edu}


\author{Robert Strausbaugh}
\affiliation{Department of Physics, Arizona State University}

\author{Nathaniel Butler}
\affiliation{School of Earth and Space Exploration, Arizona State University}

\author{William H. Lee}
\affiliation{Instituto de Astronom\'{\i}a, Universidad Nacional Aut\'onoma de M\'exico}

\author{Eleonora Troja}
\affiliation{Department of Astronomy, University of Maryland}
\affiliation{NASA Goddard Space Flight Center}

\author{Alan M. Watson}
\affiliation{Instituto de Astronom\'{\i}a, Universidad Nacional Aut\'onoma de M\'exico}

\begin{abstract}


Using deep and high-cadence gamma-ray burst (GRB) afterglow data from RATIR, we observe a sharp and achromatic light curve break 12.6 days after the GRB, accompanied by an approximately achromatic bump.  Fitting of the optical, NIR, and X-ray data suggest a very narrow (2 degree) jet which remains collimated at late-time.  We argue that the sharp light curve bump suggests an edge brightened jet, perhaps emitting only during a brief period of lateral jet expansion.   The lightcurve also exhibits a gradual spectral evolution lasting $>10$ days.  The evolution of the flux can be modeled as $\textrm{Flux} \sim \big(\frac{t}{[20 \textrm{days}]}\big)^\alpha \big(\frac{\lambda}{[800 \textrm{nm}]}\big)^\beta$, with a temporal slope $\alpha=-0.956 \pm 0.003$ and a gradually time-varying spectral slope $\beta =  (0.60 \pm 0.07)+(0.26 \pm 0.06) \textrm{log}\big(\frac{t}{20 \rm{days}}\big)$.

\end{abstract}


\section{Introduction} \label{sec:intro}
GRB 160625B was detected by NASA's Fermi Gamma-ray Space Telescope's $\gamma$-ray burst monitor \citep{gbm} as a one-second long pulse \citep{first_detect}.  Automatic follow up by the Large Area Telescope \citep{lat} resulted in detection of another bright, but longer lasting ($\approx$ 30 seconds) pulse about three minutes later.  This later pulse peaked at a visual magnitude of 7.9, and a secondary peak exhibiting significant polarization was detected 16 seconds later by the MASTER-IAC telescope \citep{master}.  We focus here on late-time, afterglow data in the $riZYJH$ bands captured with the Reionization And Transients Infra-Red/Optical camera (RATIR) \citep{ratir_first} which features spectral evolution and a sharp  bump  in  the  light  curve that were not modeled in \citep{nature_grb}.  Over fifty observing nights after the GRB, we are able to measure a so-called ``jet break'' with unprecedented cadence and sensitivity across multiple optical/NIR bands.  We also study {\it Swift}~X-ray and Ultra-Violet (UV) data captured during the same epoch.


These data potentially allow us to obtain unique constraints on the jetting of the afterglow and the possibility of lateral expansion of the jet.
At early times, the high bulk Lorentz factor, $\Gamma \approx 10^3$, of the outflow permit us to view only a narrow region of angular size $1/\Gamma$ of the jet.
The polarization detected by MASTER peaked at $8 \pm 0.5 \%$ \citep{nature_grb}, suggestive of a jet viewing angle which is slightly off-axis.
As the blast wave decelerates, more of the jet becomes visible.  Once $1/\Gamma\sim \theta_{\rm jet}$, the edge of the jet becomes visible and the flux begins declining more rapidly as the energy per solid angle begins decreasing \citep{jets,beaming,theta1/gamma}.  The edges of the jet come into causal contact at about this point, and the jet can potentially begin spreading laterally \citep[see, e.g.,][]{lateral_exp1,lateral_exp2,lateral_exp3}.  If the jet spreads, it can effectively halt the blast wave expansion and further decrease the afterglow flux \citep{beaming,jet_angle,hydro1,lateral_exp1}. 

Detailed observations and accurate models for jet breaks are critical because they allow us to determine opening angle of the jet \citep{jetangle}, which is crucial in turn for understanding GRB energetics \citep{energetics1,energetics2,jet_angle,energetics3,energetics5,gammatheta,energetics4} and rates \citep{jets,grb_rates,rates2,rates3,rates4}.  In addition, high-cadence observations with small error bars (as we have here) can potentially allow us to measure the energy and velocity structure of the jet \citep[e.g.,][]{jet_dec,jet_homo,jet_inc2,jet_unify} and to constrain the hydrodynamical processes that potentially lead to a spreading jet \citep{jet_angle,hydro1,hydro2,hydro3}.

\section{Analysis}
\label{sec:analysis}

\begin{figure}[h]
\begin{center}
\includegraphics[scale=0.7]{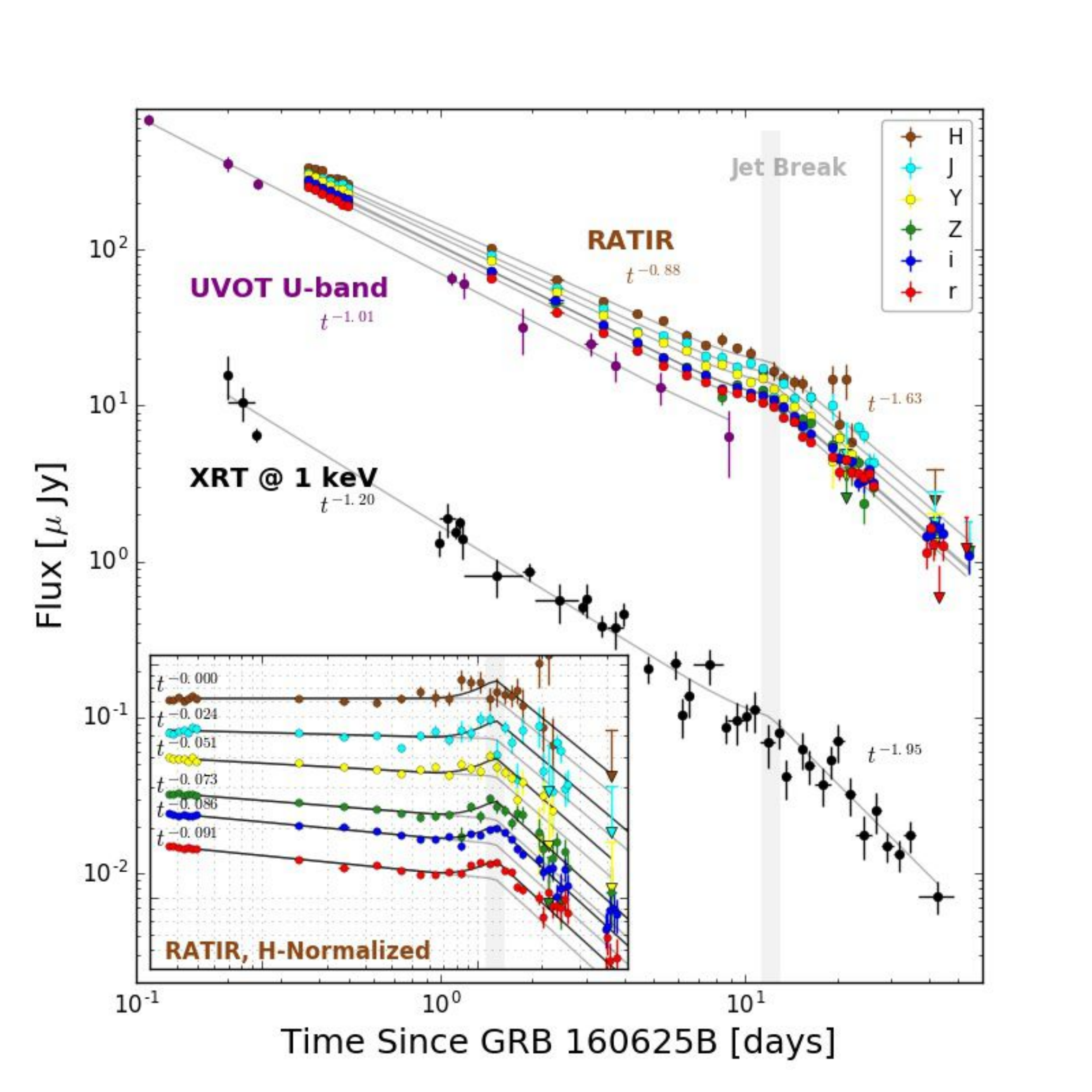}
\end{center}
\caption{The afterglow lightcurve for GRB 160625B in the $riZYJH$ bands from RATIR. X-ray and UV data are from $Swift$.  The inset lightcurves are normalized by the early time $H$-band to better display the jet break and bump.  The data in both graphs are fit with the model described in Section \ref{sec:analysis}.  Additional information about the fits can be found in Table \ref{fits}.  The data presented in this figure can be found in the Appendix.}
\label{lightcurves}
\end{figure}

RATIR photometry for GRB~160625B in the $riZYJH$ bands, reduced as described in \citet{nature_grb}, along with measurements reported by the {\it Swift}~UVOT and XRT are shown in Figure \ref{lightcurves}.
A dominant feature in the RATIR and XRT data is an apparently achromatic temporal ``jet-break" at a time of about 12 days.  Interestingly, there is a slight brightening (i.e. the temporal power-law decay is less steep around the jet break than at early times) present just prior to this jet-break.  The feature is present in all the RATIR bands with comparable amplitude, suggesting a color similar to that of the afterglow.  The jet-break, and the brief re-brightening just before it, can be seen more clearly in the inset of Figure \ref{lightcurves}, where the RATIR data have been normalized with respect to the early $H$-band behavior.

\begin{table}[h]
\caption{Light Curve Fitting Parameters}
\begin{center}

\begin{tabular}{c|cccccc} 
    Band & $\theta_1$  & $\theta_{\rm jet}$  & $B (\%)$  & $\alpha_1$ & $\alpha_2$  & $\chi^2/\nu$ \\ \hline
$r$    & $1.75 \pm 0.05$      & $2.40 \pm 0.05     $& $22.0 \pm 2.1$   & $0.971 \pm 0.002$  & $1.59 \pm 0.06$ & 1.31    \\
$i$    & $1.90 \pm 0.10$      & $2.40 \pm 0.05    $ & $18.6 \pm 1.9$   & $0.966 \pm 0.002$  & $1.64 \pm 0.05$ & 1.10    \\
$Z$    & $2.00 \pm 0.15$      & $2.35 \pm 0.05   $  & $23.4 \pm 3.6$   & $0.953 \pm 0.004$  & $1.58 \pm 0.10$ & 0.87    \\
$Y$    & $1.95 \pm 0.25$      & $2.35 \pm 0.05  $   & $17.8 \pm 4.6 $  & $0.931 \pm 0.005$  & $1.73 \pm 0.21$ & 0.80    \\
$J$    & $1.95 \pm 0.35$      & $2.35 \pm 0.15 $    & $29.8 \pm 9.8 $  & $0.904 \pm 0.005$  & $1.37 \pm 0.16$ & 2.91    \\
$H$    & $1.15 \pm 0.50$      & $2.80 \pm 1.10$     & $23.3 \pm 12.5$  & $0.880 \pm 0.006$  & $2.19 \pm 0.90$ & 1.65    \\
$UV$   & ...                & ...               & ...              & $1.013 \pm 0.032$   & ... & 0.32 \\
$X$-ray & ...               & $2.5 \pm 0.3$     & $<20.5$ (1-$\sigma$) & $1.202 \pm 0.022$ & $2.06 \pm 0.22$ & 1.64 \\\hline
\end{tabular}
\\
\vspace{0.5cm}
Fitting parameters from the solid line models plotted in Figures \ref{lightcurves} and \ref{jet_structure}, corresponding to Equation \ref{jet_model}.
\end{center}
\label{fits}
\end{table}

The Swift XRT data (Figure \ref{lightcurves}), reduced using our automated pipeline\footnote{http://butler.lab.asu.edu/swift}, show a power-law decline in flux as $t^{-1.20\pm 0.02}$ prior to the break.  The spectrum, with a mean count rate of 0.014 cps (0.3-10 keV), is well-fitted ($\chi^2/\nu = 68.57/75$) by an absorbed power-law with photon index $\Gamma=2.07 \pm 0.06$ and an absorbing column of $N_H = 4.4 \pm 0.1 \times 10^{21}$ cm$^{-2}$ at $z=1.406$ in addition to the Galactic absorbing column.  The mean unabsorbed flux is $(103 \pm 5)$ nJ at 1 keV.

\begin{figure}[h]
\begin{center}
\includegraphics[scale=0.7]{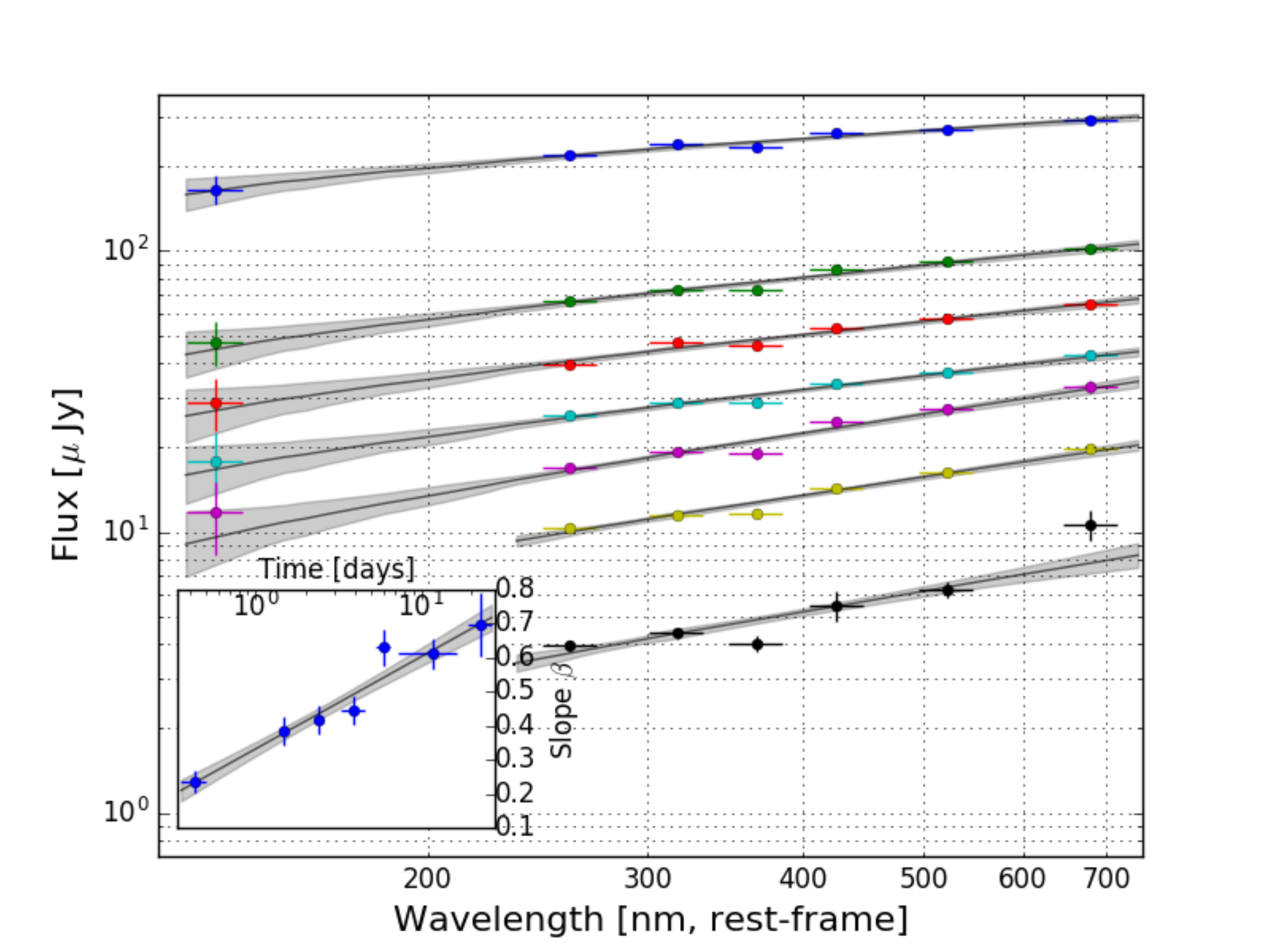}
\end{center}
\caption{The spectral evolution of GRB~160625B over the RATIR bands, as well as UV from $Swift$.  The data are fit with a power law attenuated by SMC extinction \citep{dustabs}.  The inset shows the evolution of the spectral power-law index, $\beta$, over time; the power-law index and fit statistics can be found in Table \ref{sed_fits}.  The data presented in this figure can be found in the Appendix.}
\label{sed}
\end{figure}

Assuming the standard external shock model \citep[e.g.,][]{cooling_break} for a constant density circum-burst medium (CBM), in the slow-cooling regime with a cooling break below the X-ray band, the X-ray temporal and spectral indices imply and are consistent with a power-law index for the shocked electrons of $p=2.26\pm 0.03$.  Assuming the optical/NIR bands are below the cooling break, the implied temporal decay is $t^{-0.94\pm 0.02}$.  This is similar to the typical decay laws we observe (Figure \ref{lightcurves}; Table \ref{fits}), although the observed indices are not constant across the optical/NIR bands.  The early-time optical/NIR spectral energy distribution (SED) is consistent with the expected $F_{\nu} \propto \nu^{-0.6}$ ($F_{\lambda} \propto \lambda^{0.24}$ in Figure \ref{sed}) spectrum, absorbed by $A_V \sim 0.1$ of SMC-type dust \citep{dustabs}.  The 1 keV to $r$-band flux ratio ($\sim 50$; Figure \ref{lightcurves}) is consistent with a cooling break initially near the X-ray band.

\begin{table}[h]
\caption{SED Fitting Parameters}
\begin{center}
\begin{tabular}{c|c|c}
Time (days) & $\beta$ & $\chi^2/\nu$ \\ \hline
0.36-0.51 & 0.24 $\pm$ 0.07 & 0.96 \\ 
1.41-1.52 & 0.39 $\pm$ 0.07 & 1.40 \\ 
2.27-2.52 & 0.42 $\pm$ 0.07 & 1.61 \\ 
3.30-4.53 & 0.45 $\pm$ 0.07 & 1.36 \\ 
5.31-6.45 & 0.63 $\pm$ 0.08 & 2.50 \\ 
7.30-16.41 & 0.61 $\pm$ 0.07 & 3.42 \\ 
19.25-26.51 & 0.70 $\pm$ 0.11 & 5.84 \\  \hline
\end{tabular}
\\
\vspace{-0.1cm}
\end{center}
Fits for the power-law models describing the spectral evolution of GRB~160625B plotted in Figure \ref{sed}; all models are fit using an $A_V = 0.05 \pm 0.04$ in SMC law extinction \citep{dustabs} in the host galaxy.
\label{sed_fits}
\end{table}

The temporal decay law in the optical/NIR bands flattens slightly with increasing wavelength (Figure \ref{lightcurves}, inset; Table \ref{fits}).  The data are well-fitted as  $\alpha(\lambda) = (0.938 \pm 0.003) - 2.5(0.08 \pm 0.01)\rm{log}(\lambda/{\rm [980 nm]})$.  
The result is a slow and continuous reddening that yields an optical/NIR SED (Figure \ref{sed}) described by a gradually steepening power-law index, $\beta =  (0.60 \pm 0.07)+(0.26 \pm 0.06) \textrm{log}\big(\frac{t}{20 \rm{days}}\big)$, reaching $F_{\lambda} \propto \lambda^{0.6 - 0.7}$ by the end of the observation.  The evolution of the spectral power law index -- likely due to a gradual passage of the synchrotron spectrum beginning prior to our observations -- may or may not continue through the jet-break (Figure \ref{sed}, inset).  The color transition prior to 10 days is gradual and smooth, with no break in either the spectrum or lightcurve.  We see no evidence for any strong spectral evolution during the jet break, with the synchrotron cooling frequency likely to be above the RATIR bandpass until at least approximately 30 days after the GRB.

\begin{figure}[h]
    \begin{center}
\includegraphics[scale=0.7]{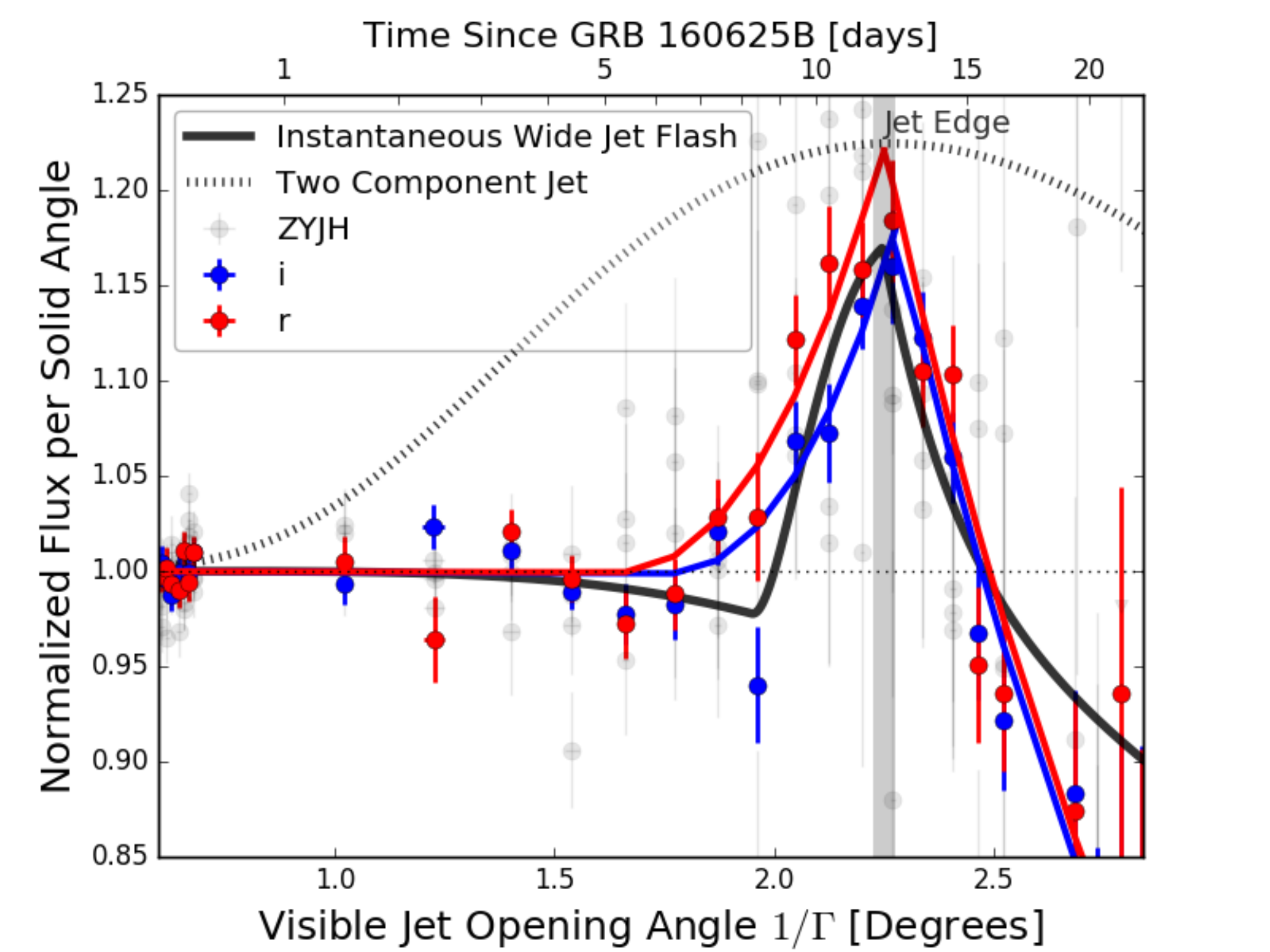}
    \end{center}
\caption{The emissivity of GRB 160625B's jet with respect to jet angle for all bands (with i and r bands highlighted), showing a structured jet with bright edges.  The blue and red curves are the model shown in Equation \ref{jet_model}; the black and gray curves show physical models derived in Section  \ref{sec:disc} for two-component jets.}
\label{jet_structure}
\end{figure}

We determine the jet opening angle, $\theta_{\rm jet} = \Gamma(t_{\rm jet})^{-1}$, using the jet break time $t_{\rm jet}$ as
\begin{equation}
    \theta_{\rm jet}=\Gamma^{-1}(t_{\rm jet})=3.27\bigg(\frac{t_{\rm{jet}}}{\rm days }\bigg)^{3/8}\bigg(\frac{1+z}{2}\bigg)^{-3/8}\bigg(\frac{E_{\rm iso}}{10^{53}~\rm{erg}}\bigg)^{-1/8}\bigg(\frac{\eta}{0.2}\bigg)^{1/8}\bigg(\frac{n}{0.1~\rm{cm^{-3}}}\bigg)^{1/8} = 2.28 \bigg(\frac{t_{\rm{jet}}}{12.6~{\rm days }}\bigg)^{3/8} {\rm degrees}
\label{jetangle}
\end{equation}
\citep{jetangle}.  Here, we have inserted values for the redshift $z$, the isotropic energy in $\gamma$-rays $E_{\rm iso}$, the efficiency of converting the ejecta kinetic energy into $\gamma$-rays $\eta$, and the CBM density $n$ from \citet{nature_grb}.  If we make the simplifying assumption that we are viewing the jet exactly on-axis, we can use Equation \ref{jetangle} to convert between observed time and the observable extent of the jet $1/\Gamma(t)$.  The light curve can then be divided by the empirical, wavelength-dependent, early-time decay law to reconstruct the apparent jet profile $F_j(\theta=1/\Gamma)$ (Figure \ref{jet_structure}).

We discuss the relation between $F_j(\theta)$ and the jet emissivity $j(\theta)$ in detail below in Section \ref{sec:disc}.  In the uniform, or homogeneous, jet model \citep[e.g.][]{jets,jet_angle}, $F_j=1$ until the edge of the jet becomes visible at $1/\Gamma =\theta_{\rm jet}$.  After this time, in the absence of jet spreading, $F_j(\theta) = (\theta_{\rm{jet}}\Gamma)^2$, and the flux steepens by a factor $(t/t_{\rm jet})^{-3/4}$ in time.  This model fits the data well at early and late time in all bands (see, Figure \ref{lightcurves}).  However, the lightcurve bump that occurs near the jet break requires an additional component.  We assume a phenomenological model:
\begin{equation}
F_j(\theta=1/\Gamma)=\begin{cases}
1, & \theta \leq \theta_1 \\

    1 + B(\theta-\theta_1)^2/(\theta_{\rm jet}-\theta_1)^2, & \theta_1 < \theta \leq \theta_{\rm jet} \\

1 + B(\theta_{\rm jet}/\theta)^2, & \theta > \theta_{\rm jet}
\end{cases}.
\label{jet_model}	
\end{equation}

The apparent jet flux $F_j(\theta)$ is constant until $1/\Gamma=\theta_1$, after which point it increases quadratically by a limb-brightening factor $B$ at the edge of the jet, $\theta_{\rm jet}$.
We find that all bands are well-fitted by such a model with consistent values for the parameters (Table \ref{fits}).  The X-ray data do not require a bump, but they also cannot rule out the optical/NIR bump at $>1 \sigma$ significance ($\Delta \chi^2=2.28$ for 2 additional degrees of freedom).  The model is also over-plotted in Figure \ref{lightcurves} using the mean fit parameters ($\theta_1 = 1.80 \pm 0.05^{\circ}$, $\theta_{\rm jet} = 2.40 \pm 0.03^{\circ}$,  $B = 20.5 \pm 1.2\%$) to compute $t^{-\alpha(\lambda)} F_j(\theta)$.

\section{Discussion}\label{sec:disc}

Bumps of varying shapes and sizes have been observed in GRB afterglows.  A contemporaneous supernova (SN) can cause a re-brightening in the afterglow lightcurves \citep{grb_sn, grb_sn2}.  However, at $z=1.406$ \citep{grb_z, grb_z2}, typical SNe (absolute magnitude $M=-19$) would be 5 magnitudes fainter than the bump in Figure \ref{lightcurves}.   The bump has a red color consistent with that of the afterglow, quite unlike the very blue color of the brightest SNe \citep[e.g.,][]{bright_sn}.  Furthermore, SNe have very broad temporal brightening features \citep[e.g.,][]{grb_sn} , very different from the sharp bump in the afterglow of GRB~160625B.  Reprocessing the afterglow light by dust in the CBM can, in principle, generate bumps in the NIR but not typically in the $r$ band \citep[e.g.,][]{dust,dust2}.  As the optical transition from reverse-shock to forward-shock dominated emission is early \citep[$t<1$ day;][]{nature_grb}, it is not likely to contribute the sharp bump 10 days after GRB~160625b. 

X-ray flaring is a common effect seen in many early afterglows \citep[e.g.,][]{grb_lc}, although no early flaring is detected in the afterglow of GRB~160625B.  Attributed to a central engine that is still active \citep{rebright_flare, grb_lc}, these features are similarly narrow in time -- $dt/t\sim 0.1$ for early \citep[e.g.,][]{chincarini07_grbflares} and late \citep[e.g.,][]{curran08_grbflares} flares -- but refreshed shocks typically occur within hours of the GRB \citep{rebright,rebright_flare} and also exhibit harder spectra than the afterglow \cite[e.g.,][]{flare_hard}.  It is important to note that there is no observed change to the color evolution in the SED around the time of the re-brightening.


It seems most natural to assume that the increase in flux just before the jet break is not coincidental, but that the phenomena are related.  However, it is important to note that the effects of relativistic beaming would permit a jet with bright edges \citep[e.g., as implied in Equation \ref{jet_model} above, or][]{jet_inc2} to be observed at quite early time, yielding smooth temporal variations in the observed flux with $dt/t\sim 1$.  A jet with a bright edge that does not change with time would produce a wide bump in the light curve starting at earlier times than the bump in Figure \ref{jet_structure}.  To see this, we can derive the observed jet structure starting with a model for the rest-frame emissivity $j^\prime$ of the jet.  The expected flux is
\begin{equation}
f_{\nu}(t) = 2 \pi D \Gamma^{-2} \int \varphi^2 d \varphi \int \frac{j_{\nu}'(t',\Omega') d \mu}{(1-\beta \mu)^2(1-\mu^2)^{3/2}}
\label{jet_model_origin}
\end{equation}
\citep[see,][]{relcorr},
where $D$ is the distance from the source to the observer and $\varphi$ is the angle to the jet edge as viewed by the observer.  Here, $\beta= v/c$ and $\Gamma=1/\sqrt{1-\beta^2}$; $\mu$ is the cosine of the angle between the velocity and the direction of the observer.  We now assume a spherical blast wave traveling directly toward the observer and a infinitesimally thin emitting shell with zero emissivity beyond an angle $\theta=\theta_{\rm jet}$:
\begin{equation}
j^\prime = A_0 t'^{-a}\nu'^{-b} \delta(r-\beta c t)H(\theta_{\rm jet}-\theta).
\label{emissivity}
\end{equation}
Here, $a$ is the power-law temporal index and $b$ is the power-law spectral index.  The rest-frame time, $t'$, and the lab-frame time, $t$, are related by  $t' = t + r \mu /c$, and $r$ is the radius of the blast wave.  The function $H$ is the Heaviside function.  Following \citet{relcorr}, we can use the delta function to integrate over the viewing angle $\varphi$ to obtain:
\begin{equation}
f_{\nu}(t) = 2\pi \beta A_0 \bigg(\frac{c }{D}\bigg)^2 \frac{t^{2-a} \nu^{-b}}{\Gamma^{2+b}} \frac{(1- \beta)^{4-a+b}}{(4-a+b)} \bigg( 1 - \bigg[\frac{(1-\beta)}{(1-\beta \mu_{min})}\bigg]^{4-a+b}\bigg),
\label{bump_start}
\end{equation}
with $\mu_{min}=\cos(\theta_{\rm jet})$.  The term in the square brackets goes to zero at early time, and the pre-factor is the flux due to a spherical, non-jetted blast wave, $f_{\nu,{\rm sphere}}$.
Defining, $F_j=f_{\nu}/f_{\nu,{\rm sphere}}$, we have:
\begin{equation}
F_j = 1 - [(1-\beta)/(1-\beta \mu_{min})]^{4-a+b} \approx 1 - (1+(\Gamma \theta_{\rm jet})^2)^{-n},
\label{ans}
\end{equation}
where we have taken the small angle limit.  Like $F_j$ above in Equation \ref{jet_model}, this function is constant ($F_j=1$) at early time and then falls like $(\Gamma \theta_{\rm jet})^2 \sim t^{-3/4}$ at late time, due to the relationship, $\Gamma \propto t^{-3/8}$, seen in Equation \ref{jetangle}.  The index $n\approx4$ affects the sharpness of the break, since the flux decays as $t^{-\alpha} \nu^{-b}$, $\alpha = 1/4+a/4+3b/8$ and $n = 5 - 4\alpha + 5b/2$.  The indices $\alpha$ and $b$ above and below the cooling break are constrained by closure relations and, in terms of the electron power law index $p$, $n = 11/2-p/2$ and $n = 7(1-p/4)$ below and above the cooling break, respectively.  Hence, for $p=2$, we expect a slightly sharper break below the cooling break ($n=4.5$) than above the cooling break ($n=3.5$).

A narrow jet ($\theta_1$) with a large $\Gamma$ enveloped by a wider jet ($\theta_2$) with a smaller $\Gamma$ can be modeled from Equation \ref{ans} as $F_j(\theta_1) + (1+B)(F_j(\theta_2)-F_j(\theta_1))$.  Plotted in Figure \ref{jet_structure} (as Two Component Jet), this model shows that relativistic beaming does not simply restrict the observer to view a portion $1/\Gamma$ of the jet.  Rather, because the emissivity versus angle is convolved with the relative Doppler factor, $1+(\Gamma \theta)^2$, to some power, a jet with an increased edge emissivity tends to produce temporally broad light curve variations ($dt/t \approx 1$). Some mechanism must be invoked to introduce additional time dependence.  A natural mechanism is the lateral spreading of the jet, which can begin around the jet break time because the entire surface of the jet is just coming into causal contact at that point.  \citet{jet_shapes} argue that the the jet angle should increase as $\theta_{jet} \approx \theta_1 + c_s/(c\Gamma)$, where $c_s$ is the sound speed, leading to an approximately constant relative Doppler factor during the expansion.  The function $F$ then remains flat for longer.  More recent work on jet expansion points towards a slower logarithmic jet expansion \citep{lateral_exp2,jet_log} as opposed to a fast exponential expansion \citep{jet_angle,hydro1,hydro2}.

To produce a narrow bump, we invoke the possibility of an instantaneous flash of emission, modeled by replacing $H$ in Equation \ref{emissivity} by $H + j_e^\prime(\theta) \delta(t'-t_1') t_1'$.  Here, $j_e^\prime(\theta)$ is a dimensionless, relative emissivity which is zero within $\theta_1$.  For $\theta>\theta_1$, we define $j_e^\prime(\theta) = B (\theta-\theta_1)^2/(\theta_{\rm jet}-\theta_1)^2$ (cf. Equation \ref{jet_model}).  With this addition, $F_j$ (Equation \ref{ans}) becomes:
\begin{equation}
F_j = 1 - [1+(\Gamma \theta_{\rm jet})^2]^{-n} + n \bigg(\frac{t}{t_1}\bigg)^{-n} j_e\bigg( \theta= \frac{1}{\Gamma} \sqrt{ \frac{t-t_1}{t_1} } \bigg),
\label{equationF1}
\end{equation}
where $t_1$ is the observer-frame time corresponding to $\theta_1$.  This model is plotted in Figure \ref{jet_structure}, with $B=26.4$\%.

Jets with either homogeneous or a brighter central region \citep{jet_dec}, viewed on-axis, are not expected to have an increase in their afterglow light curves.  Jets with a brighter central region viewed slightly off-axis, may be able to cause a brief re-brightening before the jet break.  If viewed from an angle not directly along the central axis of the jet, but still inside the jet opening angle ($0<\theta_{\rm view}<\theta_{\rm jet}$), the observer could detect an increase in flux as the brighter center of the jet came into view.  However, with these viewing conditions, we expect to see more complicated jet-break behavior on long time-scales \citep[$dt/t\sim 1$; see, e.g.,][]{jet_inc2}.  Jet models are considered in \citet{jet_inc2} which have a Gaussian energy profile and more exotic jet structures -- such as ring- or fan-shaped jets \citep{jet_shapes} -- exhibit more complex afterglow behavior (e.g. multiple jet breaks).  Two-component jets \citep{2jets_theo,2jets} create smoother bumps at earlier times (e.g. the two-component jet plotted in Figure \ref{jet_structure}), that are not consistent with our short-duration bump and the ensuing rapid steepening by $(\Gamma \theta_{\rm jet})^2 \sim t^{-3/4}$.

It is also important to note that the functional form of this steepening is inconsistent with the hypothesis of continued lateral expansion of the jet.  That expansion tends to halt the radial expansion of the fireball, producing a rapid flux decline in all bands proportional to $t^{-p}$ \citep[see,][]{jet_angle}.  We rule out that scenario at the $>4\sigma$ level (Table \ref{fits}), apparently consistent with hydrodynamical simulations \citep[e.g.,][]{jet_inc2}.
Although we think lateral expansion does not persist at late time for this afterglow, we do think it is important near the jet break time.  It is a brief period of lateral expansion lasting $dt/t\approx df/f \approx 0.2$ that allows material just outside the primary jet ($\theta>\theta_1$ in Equation \ref{jet_model}) to be shocked and to emit radiation.    Interestingly, the spectral evolution we observe for this event (Section \ref{sec:analysis}) represents a gradual loss of total blast wave energy of about 10\% as compared to canonical models involving spectral/temporal breaks \citep[e.g.,][]{cooling_break}.  It could be that this energy reservoir, lurking near the edge of the jet, is tapped to make the bump during a brief period of lateral jet expansion.

\section{Conclusion}\label{sec:conclusion}

With regular, nightly $riZYJH$ band observations over a period of weeks -- yielding a $\lesssim 3$\% typical photometric precision lightcurve -- we are able to probe the internal jet structure of the afterglow to GRB 160625B in unprecedented detail.  We observe a brief re-brightening in the afterglow light curve during the jet-break (Figure \ref{lightcurves}).  We model this increase in flux by invoking a structured jet with bright edges (Figure \ref{jet_structure}), emitting instantaneously as the the jet expands laterally for a brief period.  This interpretation is driven largely by the simultaneity of the bump and break.  The primary alternative bump explanation surviving the arguments above -- a weak pulse due to continued central engine activity -- cannot be ruled-out by the X-ray data, which do not show a clear bump but are consistent with one.  An admittedly more-pronounced X-ray bump does coincide with a probable jet break in the case of the flaring GRB~050502B \citep[e.g.,][]{other_bump1,curran08_grbflares}.  Moreover, there is at least one case \citep[e.g.,][]{other_bump2} of a similar multi-band optical bump present just before and not precisely simultaneous with a well-studied jet break.

We also observe a wavelength-dependent temporal evolution in the afterglow to GRB~160625B prior to the jet break, with temporal index $\alpha = 0.938 - 0.2 \log(\lambda/[980 ~ \rm{nm}])$.  Following the break, the temporal decay indices are consistent with those expected for a sharp-edged jet (increase by $3/4$), with no lateral expansion.


GRB~160625B exhibits a very sharply defined jet break corresponding to a very narrow jet opening angle, $\theta_{\rm jet} \approx 2^{\circ}$, indicative of nearly-on-axis viewing of a highly relativistic outflow impinging on a low density external medium \citep[see, also,][]{nature_grb}.  Typical jets should be observed at an angle $\theta_{\rm view} = \frac{2}{3} \theta_{\rm jet}$ and may or may not exhibit pronounced lateral expansion.  Both effects can introduce variations with $dt\approx t$ \citep[e.g.,][]{jet_shapes} and can tend to make jet break signatures in light curves less distinct.  Whatever mechanism created the bump for GRB~160625B (Figure \ref{lightcurves}) also contributed to making a more distinct jet break, and this effect may or may not be common.   Additional deep, high-cadence, late-time observations are required to uncover the light curve diversity and to yield a better understanding of why jet breaks are so challenging to detect and measure in the $Swift$-era \citep[e.g.,][]{swift_jetbreak}.

\section{Acknowledgements}
We thank the RATIR project team and the staff of the Observatorio Astron\'omico Nacional on Sierra San Pedro M\'artir, and acknowledge the contribution of Leonid Georgiev and Neil Gehrels to its development.  RATIR is a collaboration between the University of California, the Universidad Nacional Aut\'onoma de M\'exico, NASA Goddard Space Flight Center, and Arizona State University, benefiting from the loan of an H2RG detector and hardware and software support from Teledyne Scientific and Imaging. RATIR, the automation of the Harold L. Johnson Telescope of the Observatorio Astron\'omico Nacional on Sierra San Pedro M\'artir, and the operation of both are funded through NASA grants NNX09AH71G, NNX09AT02G, NNX10AI27G, and NNX12AE66G, CONACyT grants INFR-2009-01-122785 and CB-2008-101958 , UNAM PAPIIT grant IN113810, and UC MEXUS-CONACyT grant CN 09-283.

\appendix

\begin{longtable}{|c|c|c|c|c|c|c|c|}
\caption{RATIR GRB~160625B Data}\\
\hline
\hspace{-13 mm} \makecell{Days after \\ GRB} & \hspace{-13 mm} \makecell{Exposure \\ (minutes)} & r & i & Z & Y & J & H \\ \hline
0.37 & 1.2 & 18.24 $\pm$ 0.01 & 18.05 $\pm$ 0.01 & 17.99 $\pm$ 0.01 & 17.90 $\pm$ 0.01 & 17.79 $\pm$ 0.01 & 17.65 $\pm$ 0.02 \\ 
0.39 & 1.2 & 18.29 $\pm$ 0.01 & 18.11 $\pm$ 0.01 & 18.03 $\pm$ 0.01 & 17.96 $\pm$ 0.01 & 17.85 $\pm$ 0.01 & 17.69 $\pm$ 0.02 \\ 
0.41 & 1.2 & 18.35 $\pm$ 0.01 & 18.17 $\pm$ 0.01 & 18.08 $\pm$ 0.01 & 18.01 $\pm$ 0.01 & 17.87 $\pm$ 0.01 & 17.71 $\pm$ 0.02 \\ 
0.43 & 1.2 & 18.43 $\pm$ 0.01 & 18.23 $\pm$ 0.01 & 18.16 $\pm$ 0.01 & 18.07 $\pm$ 0.02 & 17.92 $\pm$ 0.01 & 17.82 $\pm$ 0.01 \\ 
0.45 & 1.2 & 18.46 $\pm$ 0.01 & 18.28 $\pm$ 0.01 & 18.20 $\pm$ 0.01 & 18.14 $\pm$ 0.02 & 17.99 $\pm$ 0.01 & 17.83 $\pm$ 0.01 \\ 
0.48 & 1.2 & 18.52 $\pm$ 0.01 & 18.33 $\pm$ 0.01 & 18.23 $\pm$ 0.01 & 18.15 $\pm$ 0.01 & 17.97 $\pm$ 0.01 & 17.85 $\pm$ 0.01 \\ 
0.50 & 1.2 & 18.55 $\pm$ 0.01 & 18.36 $\pm$ 0.01 & 18.30 $\pm$ 0.01 & 18.22 $\pm$ 0.01 & 18.04 $\pm$ 0.01 & 17.92 $\pm$ 0.01 \\ 
1.47 & 6.6 & 19.70 $\pm$ 0.01 & 19.51 $\pm$ 0.01 & 19.41 $\pm$ 0.02 & 19.28 $\pm$ 0.02 & 19.11 $\pm$ 0.02 & 18.96 $\pm$ 0.03 \\ 
2.39 & 15.0 & 20.26 $\pm$ 0.03 & 19.99 $\pm$ 0.01 & 19.91 $\pm$ 0.01 & 19.79 $\pm$ 0.02 & 19.62 $\pm$ 0.02 & 19.45 $\pm$ 0.02 \\ 
3.41 & 13.8 & 20.58 $\pm$ 0.01 & 20.38 $\pm$ 0.01 & 20.28 $\pm$ 0.02 & 20.15 $\pm$ 0.03 & 19.95 $\pm$ 0.03 & 19.80 $\pm$ 0.04 \\ 
4.39 & 16.8 & 20.87 $\pm$ 0.01 & 20.66 $\pm$ 0.01 & 20.56 $\pm$ 0.02 & 20.44 $\pm$ 0.03 & 20.32 $\pm$ 0.04 & 20.00 $\pm$ 0.04 \\ 
5.38 & 8.4 & 21.11 $\pm$ 0.02 & 20.89 $\pm$ 0.02 & 20.81 $\pm$ 0.04 & 20.59 $\pm$ 0.04 & 20.39 $\pm$ 0.05 & 20.12 $\pm$ 0.06 \\ 
6.39 & 8.4 & 21.27 $\pm$ 0.02 & 21.06 $\pm$ 0.02 & 20.95 $\pm$ 0.05 & 20.72 $\pm$ 0.06 & 20.50 $\pm$ 0.07 & 20.35 $\pm$ 0.09 \\ 
7.38 & 9.0 & 21.38 $\pm$ 0.02 & 21.17 $\pm$ 0.02 & 21.08 $\pm$ 0.04 & 20.96 $\pm$ 0.05 & 20.73 $\pm$ 0.06 & 20.50 $\pm$ 0.07 \\ 
8.37 & 8.4 & 21.52 $\pm$ 0.04 & 21.39 $\pm$ 0.04 & 21.43 $\pm$ 0.11 & 20.95 $\pm$ 0.08 & 20.75 $\pm$ 0.12 & 20.41 $\pm$ 0.11 \\ 
9.37 & 8.4 & 21.54 $\pm$ 0.02 & 21.37 $\pm$ 0.02 & 21.22 $\pm$ 0.05 & 21.10 $\pm$ 0.06 & 20.89 $\pm$ 0.08 & 20.55 $\pm$ 0.08 \\ 
10.36 & 6.0 & 21.61 $\pm$ 0.03 & 21.47 $\pm$ 0.03 & 21.42 $\pm$ 0.07 & 21.23 $\pm$ 0.09 & 20.83 $\pm$ 0.09 & 20.64 $\pm$ 0.11 \\ 
11.35 & 8.4 & 21.71 $\pm$ 0.02 & 21.50 $\pm$ 0.02 & 21.31 $\pm$ 0.05 & 21.15 $\pm$ 0.07 & 20.92 $\pm$ 0.08 & 20.91 $\pm$ 0.12 \\ 
12.35 & 7.8 & 21.78 $\pm$ 0.03 & 21.57 $\pm$ 0.03 & 21.48 $\pm$ 0.07 & 21.35 $\pm$ 0.10 & 21.38 $\pm$ 0.16 & 20.91 $\pm$ 0.15 \\ 
13.36 & 8.4 & 21.93 $\pm$ 0.03 & 21.69 $\pm$ 0.02 & 21.60 $\pm$ 0.06 & 21.49 $\pm$ 0.08 & 21.16 $\pm$ 0.08 & 21.02 $\pm$ 0.10 \\ 
14.41 & 14.4 & 22.01 $\pm$ 0.03 & 21.83 $\pm$ 0.02 & 21.81 $\pm$ 0.07 & 21.63 $\pm$ 0.08 & 21.40 $\pm$ 0.11 & 21.10 $\pm$ 0.12 \\ 
15.36 & 8.4 & 22.24 $\pm$ 0.05 & 22.00 $\pm$ 0.04 & 21.76 $\pm$ 0.09 & 21.92 $\pm$ 0.14 & 21.85 $\pm$ 0.18 & 21.11 $\pm$ 0.14 \\ 
16.34 & 8.4 & 22.33 $\pm$ 0.05 & 22.12 $\pm$ 0.04 & 21.83 $\pm$ 0.09 & 21.78 $\pm$ 0.14 & 21.39 $\pm$ 0.14 & 21.33 $\pm$ 0.19 \\ 
19.32 & 8.4 & 22.58 $\pm$ 0.07 & 22.34 $\pm$ 0.07 & 22.18 $\pm$ 0.15 & 22.51 $\pm$ 0.36 & 21.50 $\pm$ 0.22 & 21.04 $\pm$ 0.23 \\ 
20.32 & 7.8 & 22.82 $\pm$ 0.11 & 22.51 $\pm$ 0.09 & 22.41 $\pm$ 0.18 & 22.14 $\pm$ 0.18 & 22.03 $\pm$ 0.23 & 21.77 $\pm$ 0.24 \\ 
21.39 & 16.8 & 22.61 $\pm$ 0.13 & 22.54 $\pm$ 0.14 & $>$ 22.54 & $>$ 22.23 & $>$ 21.80 & 21.05 $\pm$ 0.28 \\ 
22.32 & 8.4 & 22.81 $\pm$ 0.12 & 22.56 $\pm$ 0.11 & 22.61 $\pm$ 0.22 & 22.39 $\pm$ 0.24 & 22.34 $\pm$ 0.31 & 22.05 $\pm$ 0.34 \\ 
23.38 & 15.6 & 22.84 $\pm$ 0.11 & 22.91 $\pm$ 0.13 & 22.48 $\pm$ 0.11 & - & 21.87 $\pm$ 0.10 & - \\ 
24.38 & 16.2 & 22.90 $\pm$ 0.16 & 22.87 $\pm$ 0.17 & 23.14 $\pm$ 0.27 & - & 21.98 $\pm$ 0.12 & - \\ 
25.39 & 18.0 & 22.85 $\pm$ 0.13 & 22.70 $\pm$ 0.13 & 22.66 $\pm$ 0.15 & - & 22.43 $\pm$ 0.18 & - \\ 
26.37 & 16.2 & 23.03 $\pm$ 0.11 & 22.91 $\pm$ 0.11 & 22.86 $\pm$ 0.15 & - & 22.43 $\pm$ 0.17 & - \\ 
39.39 & 8.4 & 24.10 $\pm$ 0.23 & 23.75 $\pm$ 0.19 & - & - & - & - \\ 
40.39 & 7.2 & 23.71 $\pm$ 0.17 & 23.75 $\pm$ 0.20 & - & - & - & - \\ 
41.39 & 7.2 & 23.98 $\pm$ 0.22 & 23.60 $\pm$ 0.17 & - & - & - & - \\ 
41.89 & 5.4 & - & - & 23.57 $\pm$ 0.32 & $>$ 23.32 & $>$ 22.89 & $>$ 22.49 \\ 
43.31 & 12.0 & $>$ 24.33 & 23.64 $\pm$ 0.21 & - & - & - & - \\ 
44.36 & 70.8 & 24.01 $\pm$ 0.22 & 23.72 $\pm$ 0.18 & - & - & - & - \\ 
52.92 & 190.2 & $>$ 23.54 & - & - & - & - & - \\ 
53.92 & 307.8 & - & 24.08 $\pm$ 0.27 & 23.86 $\pm$ 0.33 & - & $>$ 23.36 & - \\ 
\hline
\end{longtable}
\vspace{-5mm}
Magnitudes are in AB system and are not corrected for galactic extinction.

\end{document}